# Atomistic Control in Molecular Beam Epitaxy Growth of Intrinsic Magnetic Topological Insulator MnBi$_2$Te$_4$

Hyunsue Kim [1], Mengke Liu [1], Lisa Frammolino [1], Yanxing Li [1], Fan Zhang [1], Woojoo Lee [1], Chengye Dong [2], Yi-Fan Zhao [3], Guan-Yu Chen [4], Pin-Jui Hsu [4], Cui-Zu Chang [3], Joshua Robinson [2,3], Jiaqiang Yan [5], Xiaoqin Li [1], Allan H. MacDonald [1], and Chih-Kang Shih [1]

[1] *Department of Physics, The University of Texas at Austin, Austin, TX 78712, USA*
[2] *2D Crystal Consortium, The Pennsylvania State University, University Park, PA 16802, USA*
[3] *Department of Physics, The Pennsylvania State University, University Park, PA 16802, USA*
[4] *Department of Physics, National Tsing Hua University, 300044 Hsinchu, Taiwan*
[5] *Materials Science and Technology Division, Oak Ridge National Laboratory, Oak Ridge, TN 37831, USA*

**Keywords**: magnetic topological insulator, MnBi$_2$Te$_4$, molecular beam epitaxy, electronic structure, antisite defect

## Abstract

Intrinsic magnetic topological insulators have emerged as a promising platform to study the interplay between topological surface states and ferromagnetism. This unique interplay can give rise to a variety of exotic quantum phenomena, including the quantum anomalous Hall effect and axion insulating states. Here, utilizing molecular beam epitaxy (MBE), we present a comprehensive study of the growth of high-quality MnBi$_2$Te$_4$ thin films on Si (111), epitaxial graphene, and highly ordered pyrolytic graphite substrates. By combining a suite of *in-situ* characterization techniques, we obtain critical insights into the atomic-level control of MnBi$_2$Te$_4$ epitaxial growth. First, we extract the free energy landscape for the epitaxial relationship as a function of the in-plane angular distribution. Then, by employing an optimized layer-by-layer growth, we determine the chemical potential and Dirac point of the thin film at different thicknesses. Overall, these results establish a foundation for understanding the growth dynamics of MnBi$_2$Te$_4$ and pave the way for the future applications of MBE in emerging topological quantum materials.

**Main text**

The emergence of magnetic topological insulators (MTI) has created a promising material system to explore the interplay of magnetic and topological properties[1–3]. Specifically, MTI provides a rich playground to realize various topological states such as the axion insulator states and quantum anomalous Hall (QAH) states [4–7]. Early attempts to realize these exotic quantum states primarily involved the addition of ferromagnetic dopants into otherwise nonmagnetic topological insulators (TI), forming so-called extrinsic MTI (eMTI) [e.g. Cr-doped $(Bi,Sb)_2Te_3$] [5–9]. However, due to the spatial inhomogeneity of magnetic dopants, the realization of the QAH effect in eMTI requires ultra-low temperatures (30 mK). A significant recent development is the discovery of so-called intrinsic MTIs (iMTI), which have enabled the realization of the QAH effect at a much higher temperature (~2 K). Such an improvement has led to a surge of interest in these materials. The family of Mn-Bi-Te compounds, e.g. $MnBi_2Te_4$ (MBT) has been the subject of extensive study as iMTIs [9,10]. Thus far, most investigations of MBT have been carried out using bulk-grown single crystals [11–13]. However, the bulk crystal is heavily degenerate $n$-type with the Fermi level ($E_F$) located at ~0.25-0.3 eV above the Dirac point (DP), which would obscure the topological surface states due to hybridization with the bulk bands. Although successful device fabrications have been achieved on exfoliated single-crystal flakes, the method is not scalable. To overcome these limitations, developing epitaxial methods to grow iMTI thin films with precise control of thickness and effective tuning of electronic chemical potential is critical.

Molecular beam epitaxy (MBE) is a powerful technique that enables the precise control of layer thickness and dopant concentration in the growth of elemental and compound semiconductors [14,15]. Therefore, the technique is expected to be adopted to grow thin films of MBT and its variants to provide a scalable material platform. Considering that it took several decades to refine the MBE growth of III-V compounds to achieve extremely precise control, we anticipate many challenges in the MBE growth of MTIs [16,17]. Nevertheless, several groups have reported the MBE growth of MBT [18–26]. In these previous studies, the main goal was to control the growth parameters to form Mn, Bi, and Te septuple layers (SL), Te-Bi-Te-Mn-Te-Bi-Te, with minimal defects.

In this work, by combining MBE growth with *in-situ* characterization tools, we reveal several important insights into the atomistic control of MBE growth of MBT thin films. We have chosen two different substrates, Si (111) and epitaxial bilayer graphene (BL-Gr) to explore the effect of lattice match and interfacial chemical bonding. Specifically, we empirically extract the free energy landscape for the epitaxial relationship as a function of in-plane lattice orientation through spot-profile analysis of low-energy electron diffractions. Such insight provides an important guideline for the refinement of epitaxial growth. Moreover, by using *in-situ* scanning tunneling microscopy (STM), we investigate the surface structure of MBT/highly ordered pyrolytic graphite (HOPG) at the atomic scale and achieve layer-by-layer growth. By using scanning tunneling spectroscopy (STS) and angle-resolved photoelectron spectroscopy (ARPES), we investigate the evolution of the electronic chemical potential and the DP from which we extract the interplay of two key anti-site defects, $Mn_{Bi}$ as *p*-type and $Bi_{Mn}$ as *n*-type dopants. Our study lays the foundation for the future development of MBE growth of iMTI and their heterostructures.

**Results and Discussion**

**Controlling stoichiometry**

High-quality MBT films are grown using MBE on Si (111) 7×7 reconstruction surface and epitaxial BL-Gr/SiC substrates. These growths are established by the co-evaporation of Mn, Bi, and Te with an additional post-annealing process under Te ambient. Since three chemical elements are deposited on the substrate simultaneously, the parameter window for the optimal substrate temperature for self-assembly is rather narrow. At an Mn/Bi flux ratio of ~0.30, the optimal temperature range is 220-240 °C for the BL-Gr substrate and 240-260 °C for the Si substrate. During the growth, the Te flux is kept at approximately 10× the Mn flux to improve crystallinity and minimize defect formation. Further details on the growth parameters can be found in the experimental section. **Figure 1**a is a schematic showing the crystal structure of two septuple layers stacked with a van der Waals gap in between. Anti-site defects between Mn and Bi (marked as $Mn_{Bi}$ and $Bi_{Mn}$) are routinely seen in MBE-grown films, whereas the emergence of other common defects frequently observed in the bulk crystals such as $Bi_{Te}$, $Te_{Bi}$, $Te_{Mn}$, are largely suppressed.

The high-quality MBE-grown MBT films are confirmed by reflection high energy electron diffraction (RHEED) patterns shown in **Figure 1**b on Si (left) and BL-Gr (right) substrates. The bright and sharp streaks on both patterns indicate that the sample is grown uniformly with high crystallinity on both substrates. Additional streaks marked with red arrows in **Figure 1**b appear only on the MBT/BL-Gr RHEED patterns. The streaks indicate the coexistence of two different domains with 0° and 30° alignment to the substrate. Further analysis of the co-existing phases using low-energy electron diffraction (LEED) will be discussed later. **Figure 1**c shows large-scale $\theta - 2\theta$ X-ray diffraction (XRD) scans of the MBT/Si (top) and MBT/BL-Gr (bottom). Note that the substrate signals at 25-30° for Si (111) and 31.4-37° for *6H*-SiC (0006) are attenuated by a factor of 100 to prevent the substrate peak from overwhelming nearby peaks. Comparing the XRD spectra of MBT/Si and MBT/BL-Gr, the MBT film grown on the BL-Gr substrate has generally broader peaks than the film grown on the Si substrate throughout the scan. Focusing on the (009) peak, the peak width on the MBT/BL-Gr system is broader than the peak measured on the MBT/Si system by about a factor of two. This feature is most likely due to the coexistence of the two domains orientated 30° to each other. Although the films mostly comprise MBT(124) phase, a small trace of MnTe remains and appears at the shoulder of the (0024) peak, suggesting a small concentration of other phases of MBT [18]. A direct comparison in the XRD peaks with variations of flux ratio and substrate temperatures is shown in **Figure S3**.

Furthermore, core level photoelectron spectroscopy was taken on the MBT/BL-Gr system with an Al *K*α source at *hv* = 1486.6 eV (**Figure 1**d), showing the expected Mn 2p, Bi 4f, and Te 3d peaks with minimal oxidation states thanks to the quasi *in-situ* transfer between the MBE in ultra-high vacuum (UHV) to a glove box. The elemental ratio of Mn to Bi, extracted from **Figure 1**f, is 1:1.92. Comparable XPS spectra on MBT/Si and cleaved bulk MBT samples are measured under the same condition, showing the high-resolution Mn 2p, Bi 4f, and Te 3d spectra (**Figures S1** and **S2**).

**Free energy landscape for in-plane orientational alignments**

For most epitaxially grown samples, the lattice matching condition of substrates to the sample plays a critical role. Among many candidate materials for substrates, listed in **Table 1**, BT or BaF$_2$ provide minimal lattice mismatch (~1.2%) with the MBT sample. Si (111), on the other hand, has a relatively large mismatch with MBT (~11.3%). Nevertheless, Si (111) is often used as a substrate thanks to the dangling bonds at the 7×7 reconstruction surface. In **Figure 2**c, the LEED pattern of MBT grown on Si (111) 7×7 shows six-fold symmetry, reflecting a good orientational alignment. The full-width half maximum (FWHM) of the diffraction spots shows an angular dispersion of $\sigma_\theta = 11°$, presumably due to a large lattice mismatch. Similarly, the graphene $1 \times 1$ lattice has a relatively large mismatch with the MBT lattice with a lattice constant ratio of ~1.76. This large ratio, however, facilitates a 7:4 coincidental lattice matching condition (**Figure 2**a) albeit such a large integer ratio suggests that the matching might be relatively weak. On the other hand, 1.76 is close to $\sqrt{3}$, providing an alternative lattice matching condition between the MBT lattice and the graphene $\sqrt{3} \times \sqrt{3}\ R30°$ with a small mismatch of 1.6% (**Figure 2**b).

The van der Waals interface between the epitaxial BL-Gr and MBT makes it particularly interesting to investigate the competition between these two orientational alignments by analyzing the LEED patterns. In this investigation, we deposit the sample at room temperature followed by a post-annealing process. After post-annealing at 250 °C for an hour, the manifestation of the two possible orientational alignments is observed in the LEED patterns. In **Figure 2**d, along the 4:7 coincidental matching direction, the diffraction spots have a very large angular dispersion, $\sigma_\theta = 19°$, which supports the notion that the epitaxial alignment condition is not strong. On the other hand, along the Gr $\sqrt{3} \times \sqrt{3}\ R30°$ direction, the angular dispersion is less than $3.5°$, indicating a very strong epitaxial alignment. As a comparison, **Figure 2**e shows the LEED patterns for the epitaxial BL-Gr, exhibiting an angular dispersion of $1.2°$ which is near the instrumental resolution limit.

Based on the LEED study, one can construct a free energy landscape for the orientational alignment between MBT and epitaxial BL-Gr as shown in **Figure 2**f. Along the $\pm 30°, \pm 90°$, and $\pm 150°$ directions, there is a sharp and deep local free energy minimum. On the other hand, along the $0°, \pm 60°$, and $\pm 120°$ directions, the local free energy minimum is shallower but broader. The integrated counts along the Gr $1 \times 1$ and Gr $\sqrt{3} \times \sqrt{3}$ directions exhibit a relative ratio of 1.30. As this experiment starts with the thin film being deposited at room temperature, followed by post-annealing at 250 °C, one can assume that the initial nucleation sites have randomly distributed orientations. The subsequent annealing procedure drives them to local free energy minima. Therefore, our results suggest that although the local minima near 0° are shallower and broader and the local minima at 30° are deeper and narrower, the barrier between them prevents an effective re-orientation to the sharp minimum. Furthermore, using even higher annealing temperatures (nearly 300 °C) will not help since it will lead to the decomposition of MBT. We note that MBT thin films grown on both monolayer (ML)-Gr and BL-Gr substrates exhibit similar orientational alignments (**Figure S4**).

These insights suggest a compelling growth strategy. If one can precisely control the orientation of substrate steps to favor the nucleation of MBT along the Gr $\sqrt{3} \times \sqrt{3}$ direction, then the sharply defined free energy minimum will be able to drive the film growth into an epitaxial film with a sharply distributed orientation. We note that a similar strategy has been recently used to grow single-orientation transition metal dichalcogenides monolayers[27–30].

**Optimization of layer-by-layer growth**

The diffraction techniques (LEED, RHEED, and XRD) provide important information regarding crystallinity and crystal orientations. However, they do not provide microscopic information regarding the step height distributions or the distributions of various defects. For these specific observations, *in-situ* STM proves to be a valuable tool to provide important insight.

**Figure 3**a shows the STM image of an MBT film with a nominal thickness of 5.8 SL grown at a substrate temperature of 200 °C. The surface contains multiple steps with various step heights as seen in the line cut across the topography. The surface is mostly covered by fractional steps and only a small portion of the surface contains terraces with a complete quintuple layer (QL ~1 nm) or septuple layer (SL ~1.35 nm) step height. Two inset atomic images in **Figure 3**a are acquired on the terraces of a SL step height (top) and a QL height (bottom). Interestingly, the STM images resemble those obtained on cleaved $MnBi_4Te_7$ single crystals where the surface termination entails both terraces of QL height and SL height [31–33]. On the QL terminated terrace, triangular dark defects, indicative of a $Mn_{Bi}$ antisite defect, are observed. Here, a Mn atom replaces a Bi atom on the first sublayer, causing the three Te atoms to bond to it and appear in a dark depression. On the SL terminated terrace, however, it is difficult to discern specific defects. One would observe complex defect structures with an inhomogeneous background, similar to those found on the SL-terminated cleaved surface of $MnBi_4Te_7$ [31–33]. This inhomogeneity reflects a mixture of defects in different sublayers, including $Bi_{Mn}$ and $Bi_{Te}$ antisite defects. Furthermore,

the statistical distribution of step heights is extracted from an ensemble of many topography images (**Figure 3**c). The frequency distribution histogram shows a particularly large concentration of the following step heights: 0.3-0.4 nm, 0.5-0.8 nm, and 1.0-1.1 nm. 1.0-1.1 nm can seemingly be attributed to BT quintuple layers. However, fractional step heights, 0.3-0.4 nm, and 0.5-0.8 nm are likely formed due to the existence of uncombined MnTe layers as well as combinations of a few layers of MnTe, BT, and MBT, as illustrated by the cartoon images in **Figure 3**d.

By iterating the growth parameters while utilizing the result of STM investigations for feedback, we achieved the high-quality growth of pristine MBT films at a substrate temperature of 220 °C. **Figure 3**b, shows an example of the STM image of an optimized film where fractional steps are significantly minimized. The surface is left with primarily SL steps but alongside a minor presence of QL steps. Additional frequency distribution for the optimized sample is shown in **Figure S6**. This residual QL contribution indicates that the overall stoichiometry is still somewhat Mn deficient. The main difference between the samples includes a slightly elevated growth temperature by 20°C as well as a post-growth annealing process under a 33% higher Te vapor background for 2/3 of the growth time (**Figures 3**a and **3**b). Moreover, at the same Mn/Bi flux ratio, the resulting stoichiometry is subject to change when the substrate temperature is adjusted. The inset of **Figure 3**b shows a zoom-in STM atomic image revealing the Te atomic lattice on the surface where low-density $Mn_{Bi}$ antisite defects can be found.

**Evolution of Dirac point and Fermi energy**

The high-quality MBE films with controlled defect concentration provide an excellent platform to investigate the evolution of electronic structures with varying defect densities. Our recent study has shown that in the ultra-thin regime (5 SL or less), an MBT film with low defect density (< 4% $Mn_{Bi}$ antisite) exhibits a clear Dirac mass gap at low temperature (4 K) which then vanishes at 77 K (above the Neel temperature ~24 K) [34]. On the other hand, when the $Mn_{Bi}$ antisite defect density is large, the gap also vanishes. **Figure 4**a shows the STS acquired at 4 K for low $Mn_{Bi}$ defect density (~4%) on the surface of a 4 SL region where the Dirac mass gap can be observed. Also shown in the same figure are STS measurements on a 3 SL region with 9% $Mn_{Bi}$ defect where the gap completely disappears (also at 4 K). This correlation between the

Dirac mass gap and the population of $Mn_{Bi}$ defects is consistent with previous studies[34,35].

Regardless of the presence or disappearance of the Dirac mass gap, we find that the Dirac point is located at ~50 meV below the Fermi level ($E_F$) in the ultra-thin regime. This indicates that the population of the *p*-type dopant, $Mn_{Bi}$, is nearly compensated by that of the *n*-type dopant, $Bi_{Mn}$ [36]. Observation of the antisite defects using STM on the surface is relatively straightforward. However, individual $Bi_{Mn}$ antisite defects can impact an array of surface Te atoms extended across many lattice sites. Hence, quantifying the local defect concentration is nontrivial due to the interference of multiple defects. Therefore, the energy difference between $E_F$ and the DP may serve as a better indicator for the density of $Bi_{Mn}$ antisite. As the MBT/HOPG film thickness increases, we find that the sample becomes progressively more *n*-type, as observed in **Figure 4**a and summarized in **Figure 4**b. At a thickness of 10 SL, the energy difference between $E_F$ and DP reaches ~150 meV, entering the degenerate *n*-type regime. As a comparison, we show STS measurement of the UHV cleaved bulk crystal of MBT where the DP is found to be ~280 meV below the $E_F$ (**Figure 4**a).

In addition to STS studies, we also use ARPES to investigate the electronic structures of MBE films. Shown in **Figure 4**c are the ARPES spectra of several MBT films (nominally 10 SLs) grown on epitaxial BL-Gr and Si (111) substrate, respectively. These spectra are acquired at room temperature with $hv$ = 21.2 eV. It has been shown previously that at $hv$ = 21.2 eV, states near the DP are difficult to resolve, and the spectra always exhibit an apparent gap even when the surface is gapless. Nevertheless, the DP location can still be inferred from the ARPES spectra and roughly coincide with the minimum of the energy dispersion curve (EDC) at the Γ point. The ARPES results for the 10 SL MBT film grown on epitaxial BL-Gr show a DP located at -190 meV (relative to $E_F$) (**Figure 4**c), indicating that the sample at this thickness is also degenerate *n*-type. The result is close to the STS result of -150 meV acquired on a nominal 10 SL film. We also note the spectrum exhibits a lower signal-to-noise ratio in comparison to the APRES data acquired on MBT film grown on Si (111) surface. This is likely the result of the existence of two domains for MBT films grown on BL-Gr, and one of them contains a large in-plane angular distribution. Nevertheless, as the E vs k near the Γ point is nearly isotropic, identification of the DP relative to the $E_F$ is still possible. However, it should be noted that the apparent DP position is a macroscopic average, and there could be a large fluctuation in the DP position for individual grains. Moreover, in this thickness regime, quantifying thickness directly using STM is not feasible due to the loss of reference substrate in the image (**Figure S5**). Thus, the quoted thickness is estimated from the growth parameters. For a similar thickness of MBT film grown on Si (111) surfaces, the DP is located at -210 meV relative to $E_F$, also in the degenerate *n*-type regime. In comparison, the ARPES measurement of an UHV-cleaved MBT surface shows a heavily degenerate *n*-type surface with the DP located at -280 meV relative to $E_F$.

The combined STS and ARPES studies of MBE-grown MBT films on both Si (111) and epitaxial BL-Gr substrates indicate that for thick films with thickness greater than 10 SLs, the sample is degenerate *n*-type, similar to the case for the bulk MBT crystal. These results indicate that thick MBT films host large concentrations of $Bi_{Mn}$ antisite defects, in line with their bulk counterparts. On the other hand, with careful tuning, we show that in the ultra-thin regime (5 SL or less) it is possible to achieve full dopant compensation. Namely, the *n*-type $Bi_{Mn}$ dopants nearly compensate for the *p*-type $Mn_{Bi}$ dopants, resulting in an intrinsic MTI.

The evolution of the DP relative to $E_F$ as a function of film thickness highlights the interesting interplay between *n*-type $Bi_{Mn}$ and *p*-type $Mn_{Bi}$ dopants. The progressively more *n*-type characters with an increase in film thickness indicate that the concentration of $Bi_{Mn}$ defects is more than that is required to compensate for the *p*-type $Mn_{Bi}$ dopants. From this behavior, entropically mixing Mn in the Bi layer and vice versa is expected. Why does the concentration of $Bi_{Mn}$ defects positively correlate with the thickness? Previous reports have suggested that there exists a thermodynamic driving force towards the formation of high concentrations of $Bi_{Mn}$ antisites defects[36,37]. Since MnTe has a smaller lattice constant relative to that of BT, the formation of $Bi_{Mn}$ antisite defects can partially reduce the microscopic strain built up in the central Mn layer, forming an ideal Te-Bi-Te-Mn-Te-Bi-Te septuple layer. This thermodynamic driving force makes it difficult to avoid degenerate *n*-type doping in bulk MBT. In the ultra-thin regime, however, such a microscopic strain is likely to be accommodated better. Moreover, MBE growth is not a thermodynamic equilibrium process (in comparison to bulk crystal growth). Both effects may play an important role in causing nearly fully compensated doping in the ultra-thin regime. Such a conjecture could be tested in future studies and can provide a strategy to grow electronically homogeneous films over a macroscopic length scale.

**Conclusions**

In summary, we have investigated the MBE growth of MBT thin films on Si (111), BL-Gr, and HOPG substrates. The MBE growth is coupled with various *in-situ* surface characterization techniques to gain insight into the atomistic mechanism underlying the growth. By using spot profile analysis of the LEED patterns, we map out the free energy landscape for the in-plane orientational alignment between the epitaxial film and the substrate. *In-situ* STM investigations allow us to observe the film topography and defect formations. These atomic details provide timely feedback which we use to optimize the growth parameters for high-quality thin films. STS and ARPES are employed to investigate the evolution of electronic chemical potential as a function of thickness, from which we gain insight into the interplay between the *p*-type $Mn_{Bi}$ antisite defects and the *n*-type $Bi_{Mn}$ antisite defects. Our findings reveal atomistic details behind the synthesis of high-quality MTIs, providing a solid foundation for the development of thin-film devices on a larger scale. Furthermore, this significance extends beyond MBT, offering valuable insights into the engineering of other tri-elemental thin films.

**Materials and Methods**

**MBE Growth of $MnBi_2Te_4$ Films**. $MnBi_2Te_4$ thin films were grown in a home-built MBE chamber with a base pressure of ~$10^{-10}$ Torr. Highly Ordered Pyrolytic Graphite (HOPG) and bilayer epitaxial graphene substrates were outgassed at ~300 °C for 6 hours before the growth. Si (111) substrate was prepared with the standard procedure of flashing via direct current heating up to ~1200 °C for an atomically clean surface with a well-defined 7×7 reconstruction. High-purity Mn (99.99%), Bi (99.999%), and Te (99.999%) were co-evaporated from standard Knudsen cells. Samples studied with STM and ARPES are grown at 240 °C (on HOPG, graphene), 225 °C (on Si), and post-annealed at the growth temperature in an excess Te ambient. Samples studied with LEED and RHEED were deposited at room temperature (~300 K) and post-annealed at 250°C in an excess Te ambient.

**In-Situ STM/STS, ARPES Measurements.** Using a home-built UHV transfer system with base pressure ~$10^{-10}$ Torr to maintain the cleanness of the film, samples were transferred from the MBE chamber to STM (base pressure ~$10^{-11}$ Torr) and ARPES (base pressure ~$10^{-11}$ Torr) chambers without any exposure to air. STM/S measurements were conducted at 4.3 K, 77 K. The W tip was prepared by electrochemical etching and cleaned by in situ electron-beam heating. STS dI/dV spectra were measured using a standard lock-in technique with feedback off, whose modulation frequency is 490Hz. The ARPES measurements were carried out using a helium lamp with a beam spot diameter of ~300 μm, using a Scienta R3000 electron energy analyzer. The measurements were done with He I (~21.2eV) at room temperature.

**Figures**

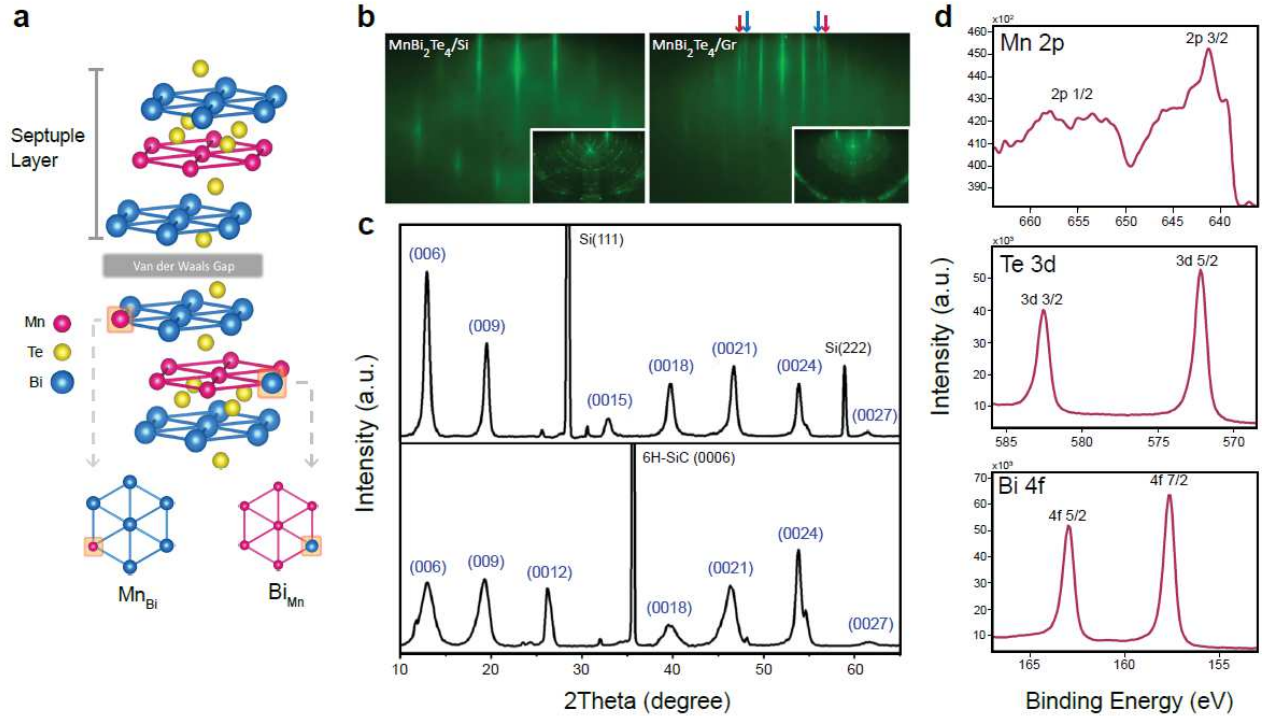

**Figure 1.** Crystal growth of a few MnBi$_2$Te$_4$ septuple layers. (a) Schematic of the crystal structure of the MnBi$_2$Te$_4$ septuple layer with anti-site defects on Mn and Bi sites, which are common defects seen in MBE-grown samples. There are other defects identified in bulk Bi$_{Te}$, Te$_{Bi,Mn}$, etc. (b) RHEED images of a MBT/Si sample (left) and substrate (inset) and RHEED images of MBT/BL-Gr sample (right) and substrate (inset). (c) XRD on MBT on Si (top) and BL-Gr (bottom), indicating both samples primarily consist of MBT (124). (d) XPS spectra on MBT/BL-Gr showing high-resolution Mn 2p, Bi 4f, and Te 3d spectra.

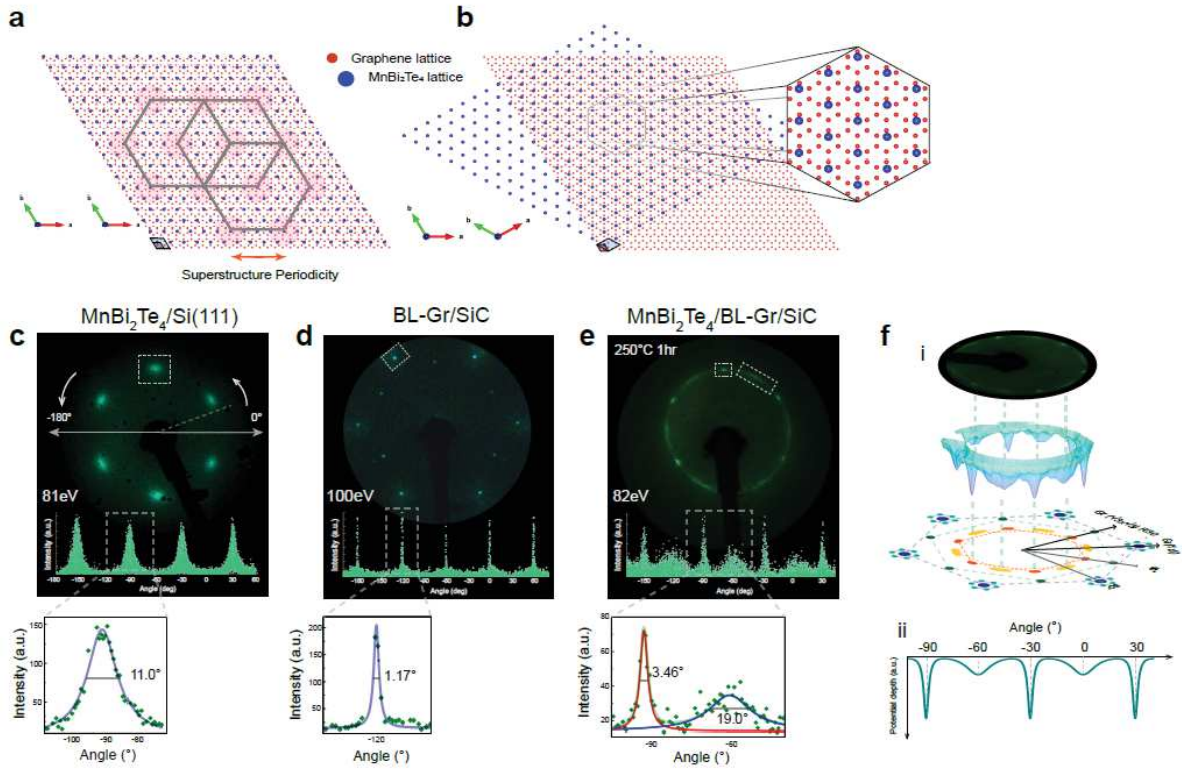

**Figure 2.** Sample – Substrate alignment MnBi$_2$Te$_4$ on Si (111) and epitaxial Gr. (a, b) superstructures between MBT (124) and epitaxial graphene substrate at 0° and 30° relative rotations, respectively. (c-e) LEED patterns of MBT/Si at electron energy 81 eV, and MBT/BL-Gr annealed at 250 °C for 1 hour, measured at $hv$ = 82 eV, bilayer graphene substrate measured at $hv$ = 100 eV respectively. Corresponding radial intensity cut along the direction shown in the panel. FWHM of selected peaks in each radial cut. The selected peaks are marked with a white dashed box in LEED patterns. (f) Qualitative sketch of the potential energy landscape obtained by inverting the LEED spot profile intensity as a function of the rotation angle.

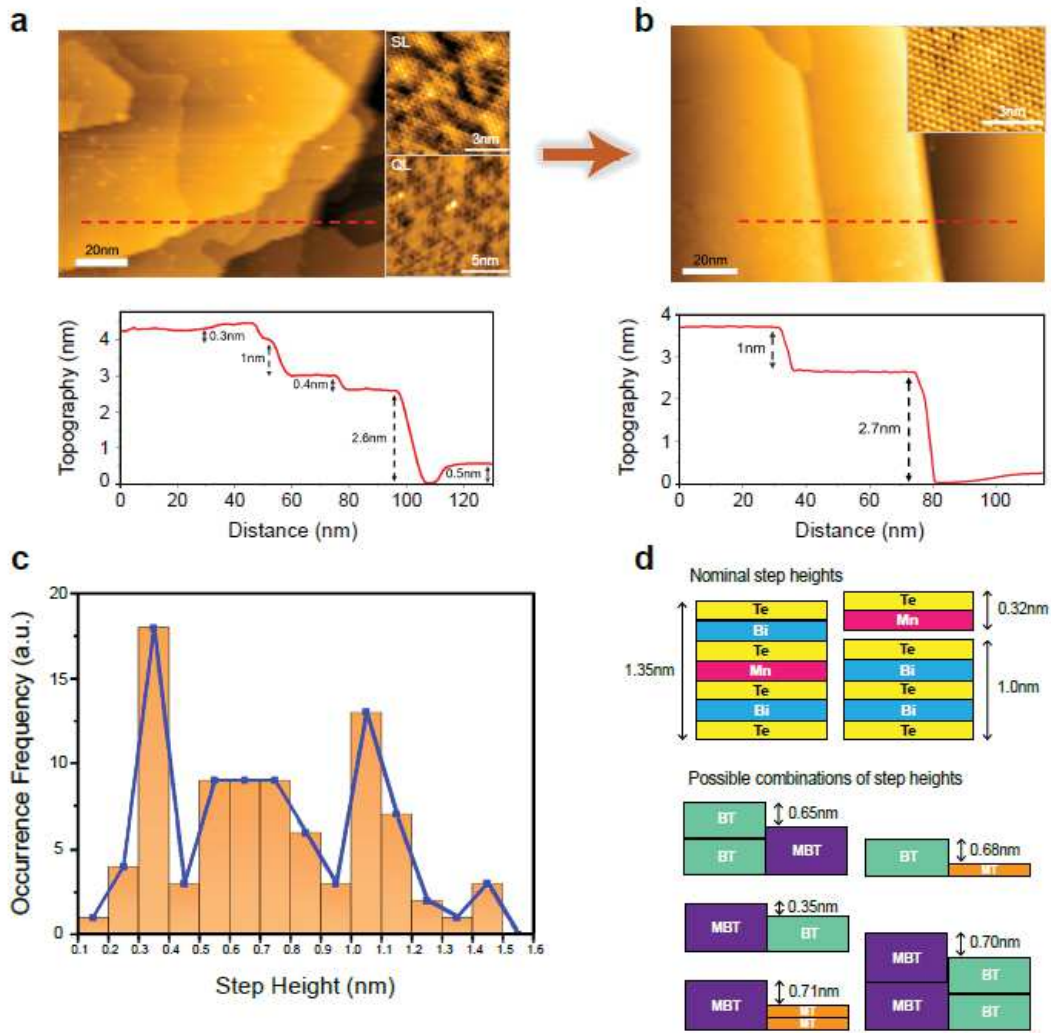

**Figure 3.** STM, fractional step analysis. (a) STM topography ($V_{bias}$= 4 V, $I_{tun.}$= 5 pA) of MBT/HOPG sample annealed at 200 °C. Corresponding profile cut entailing fractional step heights largely distributed around 0.3 nm and 0.7 nm. Atomic images of SL ($V_{bias}$= 1 V, $I_{tun.}$= 20 pA) and QL ($V_{bias}$= 1 V, $I_{tun.}$= 25 pA) terminated surfaces with signature defects. (b) STM topography ($V_{bias}$= 2.5 V, $I_{tun.}$= 15 pA) of MBT/HOPG sample annealed at 220 °C. Corresponding profile cutting through only containing mostly SL and QL. Inset represents SL termination atomic image ($V_{bias}$= -0.8 V, $I_{tun.}$= -25 pA). (c) Occurrence frequency of heights varying from 0.1 to 1.5nm. (d) Cartoon showing possible combinations of $MnBi_2Te_4$, $Bi_2Te_3$, and MnTe.

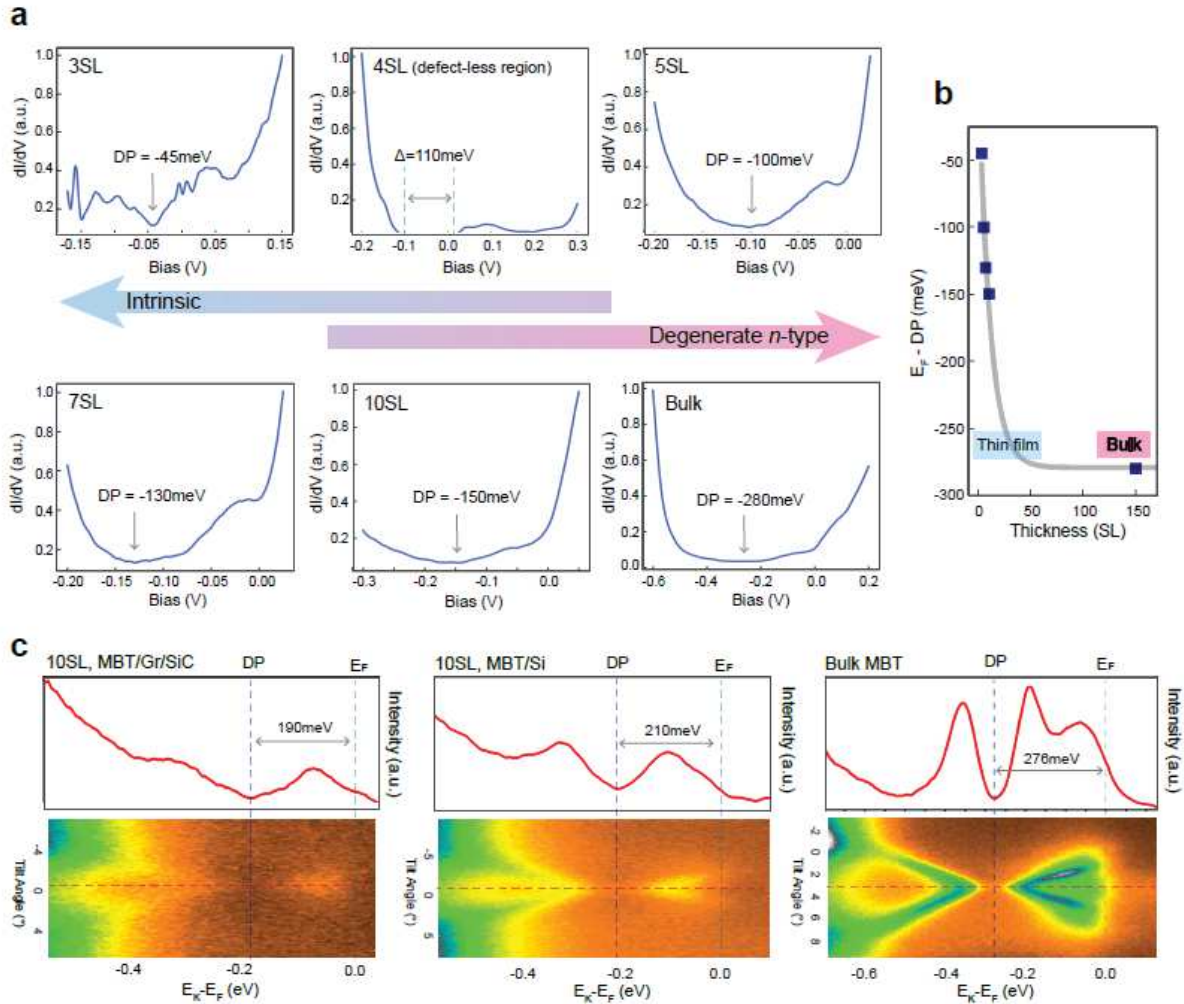

**Figure 4.** Thickness and defect concentration dependent DP evolution relative to $E_F$. (a) Shift in Dirac point relative to $E_F$ can be observed in d$I$/d$V$ spectra on 3 SL, 5 SL, 7 SL, 10 SL MBT/HOPG, and bulk MBT single crystal. The pristine area on 4 SL MBT/HOPG thin film shows a gap ~110 meV. (b) A plot of $E_F$ – DP extracted from STS as a function of thickness. (c) Evolution of DP as thickness increases. ARPES spectra on 10 SL MBT/BL-Gr, 10SL MBT/Si, and Bulk MBT in accord with STS results.

| System | $a_{cubic}$ | $a_{basal}$ | $\sqrt{3} \times \sqrt{3}\ R30°$ | Lattice Match |
|---|---|---|---|---|
| MnBi$_2$Te$_4$ | | 4.33 | | 1.00 |
| Bi$_2$Te$_3$ (0001) | | 4.38 | | 1.012 |
| BaF$_2$ (111) | 6.20 | 4.38 | | 1.012 |
| GaAs (111) | 5.65 | 4.00 | | 0.923 |
| Si (111) | 5.43 | 3.84 | | 0.887 |
| STO (111) | 3.91 | 2.76 | | (3:2) 0.956 |
| Al$_2$O$_3$ (0001) | | 4.79 | | 1.105 |
| Graphene | | 2.46 | 4.26 | (7:4) 0.994; ($\sqrt{3}$) 0.984 |

**Table 1.** Substrate lattice matching condition compared to MnBi$_2$Te$_4$ lattice.

**Acknowledgment**

This work is primarily supported by the NSF through the Center for Dynamics and Control of Materials: an NSF Materials Research Science and Engineering Center under cooperative agreement no. DMR-1720595 and the US Air Force grant no. FA2386-21-1-4061. Other support is from the Air Force Office of Scientific Research grant no. FA2386-21-1-4067. Work done in Penn State is supported by 2DCC-MIP under NSF cooperative agreement DMR-1539916, DMR-2039351, ARO Award (W911NF2210159), as well as the support from Gordon and Betty Moore Foundation's EPiQS Initiative (GBMF9063 to C.-Z. C.). Work at Oak Ridge National Laboratory was supported by the US Department of Energy, Office of Science, Basic Energy Sciences, Materials Sciences and Engineering Division. P.J.H. acknowledges support from the National Science and Technology Council of Taiwan under Grant No. NSTC-112-2636-M-007-006 and Grant No. NSTC-110-2124-M-A49-008-MY3.